\documentstyle[11pt,doublespace]{article}

\def\psn{\par\smallskip\noindent}
\def\E{{\bf E}}
\def\H{{\bf H}}
\def\P{{\bf P}}

\begin{document}

\title{\bf Lossless Tapers, Gaussian Beams, Free-Space Modes: Standing
Waves Versus Through-Flowing Waves}

\author{Antonio-D. Capobianco, Michele Midrio,
Carlo G. Someda
and Stefano Curtarolo\thanks{Present address: Dipartimento di Fisica 
G. Galilei, Universit\'a degli Studi di Padova.} \\
Istituto Nazionale per la Fisica della Materia \\
Dipartimento di Elettronica e Informatica \\
Universit\`a degli Studi di Padova \\Via G. Gradenigo 6/A
35131 Padova, Italy.}

\date{}                  

\maketitle

\begin{abstract}
It was noticed in the past that, to avoid physical inconsistencies,
in Marcatili's lossless tapers through-flowing waves must be
drastically different from standing waves. First, we reconfirm this
by means of numerical results based on an extended BPM algorithm. 
Next, we show that this apparently surprising behavior is a
straightforward fallout of Maxwell's equations. Very similar 
remarks apply to Gaussian beams in a homogeneous medium. As a
consequence, Gaussian beams are shown to carry reactive powers, and 
their active power distributions depart slightly from their
standard pictures. Similar conclusions hold for free-space
modes expressed in terms of Bessel functions.
\end{abstract}

\newpage

\section{Introduction.}

In 1985, Marcatili  infringed a historical taboo, showing
that one can conceive and design, at least on paper, dielectric tapers 
and bent waveguides that are strictly lossless \cite{Mar85}. 
The key feature, shared 
by the infinitely many structures which obey Marcatili's recipe, is that
the phase fronts of their guided modes are
closed surfaces. Phase fronts which extend to infinity
in a direction orthogonal to that of propagation entail radiation
loss: closed fronts can overcome this problem. Shortly later, however,
it was pointed out that this recipe could generate
some incosistencies \cite{MaSo87}. In fact, a traveling wave with a 
closed phase
front is either exploding from a point, or a line, or a 
localized surface, or collapsing onto such a set. In a lossless medium 
and in the absence of sources or sinks, this picture is untenable. 
On the other hand, it was also pointed
out in ref.\cite{MaSo87} that a standing wave with closed constant-amplitude
surfaces is perfectly meaningful under a physical viewpoint. Therefore, 
a through-flowing wave through any of Marcatili's lossless tapers or bends
has to be described in this way: the incoming wave must be decomposed
as the sum of two standing waves, of opposite parity with respect to
a suitable symmetry plane (or, more generally, surface). The output 
wave is then found as the sum of the values taken by the two standing 
waves at the other extremity of the device. Another point raised 
in ref.\cite{MaSo87} was that very similar remarks apply to Gaussian beams 
in free space. If applied literally to a traveling wave, the standard 
mathematics which is found in any textbook on Gaussian beams would 
entail that such beams either explode from their waist plane or 
implode from it: once again, a physically meaningless picture.

Later on, the literature showed that these problems were not dealt with 
for a long time. Recently, though, we observed several symptoms
of a renewed interest in low-loss \cite{MKM94, Man95, Vas94} and lossless
\cite{Wu96, Lee97} tapers or bends. This induced us to try to go beyond the
results of ref.\cite{MaSo87}, aiming at clarifying more deeply the 
difference between through-flowing and standing waves in Marcatili's 
tapers and in Gaussian beams. Our new results, reported in this paper, 
can be summarized as follows. In Section 2, we
show that a numerical analysis (based on an extended BPM algorithm)
of Marcatili's tapers reconfirms that indeed through-flowing waves are
drastically different from standing waves. The latter ones match very
well the analytical predictions of the original recipe 
given in ref.\cite{Mar85}, but
through-flowing waves have open wave fronts, so that they do not entail any
physical paradox. In Section 3, we provide an analytical discussion
of why, in contrast to classical cases like plane waves
in a homogeneous medium or guided modes in longitudinally
invariant waveguides, through-flowing waves in Marcatili's tapers 
are so different from standing waves. We show that the difference 
is a straightforward fallout of Maxwell's equations. Although this 
entails that through-flowing waves in Marcatili's tapers are
never strictly lossless, nonetheless our numerical results reconfirm
that the recipes given in ref.\cite{Mar85} do yield extremely low radiation
losses. In Section 4 we address the very similar problem of
Gaussian beams in a homogeneous medium. We show that physical
inconsistencies affecting the naive picture of a traveling Gaussian beam 
disappear, as soon as Maxwell's equations are handled with sufficient care. 
In Section 5, we focus our attention on those waves which were
identified in ref.\cite{MaSo87} as the true free-space modes (as 
opposed to Gaussian beams). Also for these we show that the picture 
of a through-flowing wave is affected in some of its significant
features, if Maxwell's equtions are used to go a few 
steps further than what had been done in ref.\cite{MaSo87}.

The common key feature shared by Marcatili's tapers, Gaussian
beams, and free-space modes, is that in general the Poynting vector
is not simply proportional to the square 
of the modulus of the electric field. The actual
power distribution in space is more complicated. In particular,
in contrast to classical cases, through-flowing fields in these 
problems are always characterized by nonvanishing reactive powers, which 
reach their highest levels in the proximity of the waist of the taper
or beam. This indicates that through-flowing waves do not have all the
features of a {\it pure} traveling wave, which, by definition, has a
standing-wave ratio identical to $1$, and thus cannot carry any 
reactive power.

\section{Marcatili's tapers: numerical results.}

The geometry of Marcatili's tapers can possibly be very complicated
(e.g., see ref.\cite{SaMa-1991}). Here, however, we prefer to adopt
a simple shape, to avoid that geometrical features may blur
the basic physics we were trying to clarify. The results on which we 
focus in this Section refer to a single-mode taper whose graded-index 
core region is delimited by the two branches of a hyperbola (labels 
A and A' in Figs.(1) and (2)), and has a mirror symmetry with respect to 
its waist plane. According to the terminology of ref.\cite{Mar85}, 
this is a ``superlinear'' taper, with an index distribution
                                          
\begin{equation}
n=\left\{
\begin{tabular}{ll}
$n_0\sqrt{1+2\Delta/(\cosh^2\eta-\sin^2\theta)}$ & for $\theta_1<\theta<\theta_2$ \\
$n_0$ & for $\theta_1>\theta>\theta_2$
\end{tabular}
\right.
\end{equation}

where $\eta$ and $\vartheta$ are the elliptical coordinates, in the plane
of Figs.(1) and (2). Fig.(1) refers to standing waves, of even (part a) and
odd (part b) symmetry with respect to the waist plane. The closed lines are 
constant-amplitude contour plots. They are essentially elliptical,
so they agree very well with the predictions of ref.\cite{Mar85}.

As mentioned briefly in the Introduction, these results were generated using 
an extended BPM, which deserves a short description. It is
well known that standard BPM codes are suitable to track only
traveling waves, as they do not account for backward waves. Our code 
(which uses a Pade's operator of order (5,5)) generates a traveling wave, 
but the direction of propagation is reversed
whenever the wave reaches one of the taper ends. In order to
generate a single-mode standing wave, each reflection should take
place on a surface whose shape matches exactly that of the wave front. 
This is very difficult to implement numerically, especially as
long as the wave front shape is the unknownfeature one is looking for. 
But the problem can be circumvented,
by letting each reflection take place on a {\it phase-conjugation} flat
mirror. Our code adopts this solution, and calculates then, at each 
point in the taper, the sum of the forward and backward fields. The 
process stops when the difference
between two consecutive iterations is below a given threshold.

Fig.(2) refers to a through-flowing wave. The almost vertical 
dark lines
in  part a) are the phase fronts. They are drastically different from those
predicted by the analytical theory of ref.\cite{Mar85}, which are
exemplified in the same figure as a set of confocal ellipses. Note that
the through-flowing wave has been fabricated numerically in two ways. 
One was simply
to launch a suitable transverse field distribution, and track it down the
taper, with a standard BPM code. The other one was to calculate the linear
combination (with coefficients $1$ and $j$) of the even and odd standing 
waves shown in Fig.(1). The results obtained in these two ways are
undistinguishable one from the other. This, altogether, 
proves that indeed through-flowing
waves are drastically different from standing ones. In particular,
as we said in the Introduction, through-flowing waves are totally free 
from any untenable feature under the viewpoint of energy conservation.

Fig.(2.b) shows a field amplitude contour plot for the same
through-flowing wave as in Fig.(2.a). It indicates that, in spite of all 
the matters of principle which make a through-flowing wave different from
a standing one, its propagation through the
taper is indeed almost adiabatic.
Therefore, as anticipated in the Introduction, insertion losses of
Marcatili's tapers are very low, at least as long as the length to
width ratio is not too small, although not strictly zero. A
typical example is shown in Fig.(3). It refers to a 
taper like that of Figs.(1) and (2), whose total length is $2.5 \mu m$, 
whose waist width is $0.55 \mu m$, and whose initial (and final) 
width is $1.65 \mu m$. The BPM calculations yield a lost power fraction
of $1.4 \times 10^{-4}$ at a wavelength of $1.55 \mu m$.

\section{Marcatili's tapers: analytical remarks.}

For the sake of clarity, let us restrict ourselves to the case of
two-dimensional tapers, like those of the previous section, where
the geometry and the index distribution are independent of the
$z$~coordinate, orthogonal to the plane of the figures. However,
our remarks will apply to 3-D structures also.

The index distributions that were identified in ref.\cite{Mar85} are such
that the TE modes (electric field parallel to $z$) satisfy {\it rigorously} 
a wave equation which can be solved by separation of variables. Obviously,
the same equation is satisfied rigorously by the transverse component of
the magnetic field, as well. However, in general, if we take 
two identical solutions of these two wave equations (except for a 
proportionality constant), it is easy to verify that they {\ do not 
satisfy Maxwell's equations}. This statement could be tested, 
for example, on the superlinear taper of the previous Section. 
However, this proof would be mathematically cumbersome,
requiring use of Mathieu functions of the fourth kind, which satisfy the
wave equation in the elliptic coordinate system. A much simpler,
yet enlightening example, is the device which was referred to 
in ref.\cite{Mar85} as ``linear taper'' : a wedged-shape region, with
a suitable index distribution, where only one guided mode can propagate 
in the radial direction. It is perfectly legitimate to say that 
the dependence of $\E_z$ on the radial coordinate is expressed by 
a Hankel function, whose imaginary order, $i\nu$, is related to 
the features of the individual taper \cite{Mar85}. 
What one cannot extrapolate from this, 
is that the same holds for the magnetic field. In fact, let us calculate 
the curl of the electric field. We find that the azimuthal component of 
the magnetic field is proportional to the {\it first derivative} of 
the Hankel function with respect to its argument (proportional to the
radial coordinate). This derivative is never proportional to the 
function itself. This is a drastic difference with respect to
plane waves, or to guided modes of longitudinally invariant waveguides, 
where the derivative of an exponential function (expressing the 
dependency on the longitudinal coordinate) remains proportional to 
the function itself. We see that in Marcatili's tapers, although 
$\E_z$ and $\H_{\vartheta}$, on any wavefront, have identical 
dependencies on the transverse coordinates, nevertheless it is 
troublesome to define a wave impedance, because they do not vary
in identical fashions along the coordinate of propagation.
It is equally risky to derive claims \cite{Mar85} regarding the Poynting
vector from just the spatial distribution of the electric field,
skipping the details of the magnetic field.

Let us strengthen our point with a few calculations, which aim at 
proving explicitly that a TE wave, whose radial
dependence is expressed by a Hankel function of imaginary order, $H_{i\nu}$,
cannot be a pure {\it traveling} wave along a linear taper. 
As we just said, if $\E_z$ is proportional to $H_{i\nu}$, 
then Maxwell's equations say that $\H_\vartheta$ is proportional to
$iH'_{i\nu}$. The radial component of the Poynting vector is
proportional to $iH_{i\nu} (H'_{i\nu})^*$. In a purely traveling wave,
by definition there is no reactive power flowing in the 
direction of propagation. In the case at hand, this would imply
$|H_{i\nu}|^2=constant$ along the radial direction, a requirement 
which cannot be satisfied by Hankel functions. Incidentally, note
that exponential functions, which describe traveling plane waves and 
modes of longitudinally invariant waveguides, do satisfy this type
of requirement. Coming back to the linear taper, it is easy to show
that the requirement which chararcterizes a purely standing 
wave - zero active power in the radial direction - is satisfied
by Bessel functions of order $i\nu$. This reconfirms that the exact 
modes of the lossless tapers found in ref.\cite{Mar85} are only standing waves.
{\it Any} through-flowing wave along such a taper must be expressed as
a linear combination of such standing waves. This point had already been
stated in ref.\cite{MaSo87}; the new contribution of this paper consists
in showing that the impossibility of expressing a traveling wave
in terms of one simple function of the coordinate along which 
it propagates (say, the Hankel function in the previous example)
is a consequence not of singularities, but of the inherent nature
of partially standing wave which characterizes such functions.

\section{Analysis of Gaussian beams.}

One of the points raised in ref.\cite{MaSo87} was that the usual
picture of a Gaussian beam may run into the same physical 
incosistency as a single traveling wave in a Marcatili taper.
In fact, if we read the mathematics of Gaussian beams  
in a literal way, we find that the phase fronts are ellipsoidal
surfaces, which are closed surfaces. A traveling wave with a
closed phase front is either exploding from a localized source,
or imploding onto a localized absorber. In a lossless homogeneous medium,
without sources, this is untenable. For a loosely focused beam, whose phase
fronts are almost flat, the field amplitude is negligibly small
in the regions -~far from the beam axis~- where the two ``halves'' of
a phase front (one half on each side of the beam waist) meet. Consequently,
the point that the phase fronts are closed surfaces appears not to be,
in such a case, of practical relevance. On the contrary,
in a tightly focused beam this fact is not irrelevant, and could
explain some discrepancies between experiments and the simplest theories,
which have been observed and reported in the literature. Once again,
as noted in ref.\cite{MaSo87}, a standing wave with
closed constant-amplitude surfaces is physically meaningful, regardless
of how tightly it is focused. Henceforth, a traveling Gaussian beam
which passes through its waist plane can be modeled correctly as
the sum of two standing waves, of opposite parity with respect to
the beam waist plane. 

In this Section, we will show that the difference between traveling 
and standing waves - a deep difference as a matter of principle - can 
be explained as in the previous Section, using Maxwell's equations. 
Furthermore, this procedure will enable us to find quantitative 
criteria to assess when these changes with respect to the
naive theory (where the Poynting vector is simply taken to be 
proportional to the square of the electric field) become of practical 
relevance. We will also show that the signs of the terms which are
usually neglected in the power distribution depend on the beam 
polarization. To underline this, we speak of transverse-electric (TE) 
and transverse-magnetic (TM)
Gaussian beams, in contrast to the classical $TEM_{m,n}$ terminology.
We will deal explicitly only with the TE case. The reader may easily
derive the TM case by duality. 

In the paraxial approximation around
the $z$-axis, $\left|\frac{\partial^2\phi}{\partial z^2}\right|
\ll 2k\left|\frac{\partial\phi}{\partial z}\right|$, the scalar
Helmholtz equation, in an indefinitely extended homogeneous medium, becomes
\begin{equation}
\frac{\partial^2\phi}{\partial x^2} + \frac{\partial^2\phi}{\partial y^2}
= 2jk\frac{\partial\phi}{\partial z}
\label{eq:Helmholtz}
\end{equation}
where $\E=\phi(x,y,z)\exp(-jkz)$. The features which we want to outline
can be extracted from any solution of (\ref{eq:Helmholtz}).
So, let us focus on the
simplest one, the so-called $TEM_{00}$ mode, namely
\begin{equation}
\E=\E_0 e^{-jkz} e^{-jP} e^{-j\frac{k}{2q}r^2}
\label{eq:ex}
\end{equation}
where
\begin{equation}
\frac{1}{q(z)}=\frac{1}{R(z)}-2j\frac{1}{kw^2(z)} \quad,
\quad P(z)=-j\,{\rm ln}\left[1+{z\over q_0}\right]
\end{equation}
$R$ being the radius of curvature of the beam phase front, and $w$ the beam
width at the 1/e amplitude level. Their $z$-dependence can be found in
any textbook (e.g., ref.\cite{Yar76}, Sect. 3.2).

It is elementary, starting from Maxwell's equations, to show that in the
paraxial approximation {\it all} the components of the electric field
vector, {\it and all} those of the magnetic field vector, satisfy
(\ref{eq:Helmholtz}).
What is usually taken for granted is that, for a
beam whose transverse electric field, say $\E_x$, is expressed by (\ref{eq:ex}),
the transverse magnetic field $\H_y$ is also of the form (\ref{eq:ex}), so that 
the wave impedance $\E_x/\H_y$ is constant in space.In reality, this is 
not true: if we insert $\E_x$
expressed by (\ref{eq:ex}) into Maxwell's equation $\nabla\times\E =
-j\omega\mu\H$, we get
\begin{equation}
\H_y = \frac{\E_x}{-j\omega\mu} S
\end{equation}
where
\begin{equation}
S = -jk+2j\frac{1}{kw^2}-jk\left[\frac{2r^2}{k^2w^4}-\frac{r^2}{2R^2}\right]
+\left[\frac{2r^2}{Rw^2}-\frac{1}{R}\right]
\label{eq:S}
\end{equation}
For $\H_y$ to be proportional to $\E_x$ and the wave impedance to be
equal to $\eta=\sqrt{\mu/\epsilon}$, this expression should reduce to
its first term. The second term, small and independent of the 
transverse coordinates, is insignificant. The following terms are 
small compared to $k$, but not negligible, not even in the
paraxial approximation, as the reader can check by calculating the second
derivative with respect to $z$ and comparing it to (\ref{eq:S})
multiplied by $k$.

What can make these terms important (at least as a matter of principle)
is that they depend on the transverse
coordinates. The third term in (\ref{eq:S}) affects the real part of
the Poynting vector, so that the active power density 
flowing through any cross section of the beam is not simply 
$|\E_x|^2/(2\eta)$. The last term gives
rise to an imaginary component of the Poynting vector, indicating
that there is reactive power present in the beam - a feature
in contrast with the naive model of the beam as a pure traveling wave. 
Let us look at both terms in more detail.

The third term is the sum of two quantities of opposite sign, with
identical dependence on the transverse coordinates through $r^2=x^2+y^2$,
but different dependences on $z$ through $1/R$ and $1/w^2$. Their sum cancels
out exactly for $z=\pm \pi w_0^2/\lambda=z_R$, i.e. at the two extremes of
the so-called Rayleigh range ($w_0$ being the spot size at the beam waist).
On those two planes, but only there, the wave impedance is independent 
of $r$, and equal to $\eta$. Within the Rayleigh range, the positive
contribution dominates, and the wave impedance (equal to $\eta$ only for $r=0$)
decreases as the distance from the $z$-axis grows. For example, on the
waist plane, where $1/R=0$, at $r=0.3 z_R$ (where the field is still
appreciable, in a very tightly focused beam) the wave impedance is about
$5 \%$ smaller than $\eta$. On the contrary, out of the Rayleigh range
the sum in question becomes negative. Its magnitude reaches a maximum
at a distance $z=\pm \sqrt{3} z_R$, and then decays to zero as $z$
tends to infinity. The corresponding corrections on the wave impedance
are negligible where the field amplitude is significant. Therefore, we 
may conclude that the usual TEM model
is perfectly adequate out of the Rayleigh range.

The last term in (\ref{eq:S}), giving rise to reactive power, also consists
of two contributions of opposite sign, whose sum cancels out along
the hyperboloid $r^2=w^2/2$ and has odd parity with respect to 
the waist plane, $z=0$. Poynting's theorem, in a lossless medium
and with no sources, leads then to the conclusion that electric and
magnetic energy densities are not equally stored, at all points, 
in a Gaussian beam.
In the TE case, there is more electric than magnetic energy stored
near the beam axis, and this becomes more evident as one gets
closer to the waist. The reverse is true in the periphery. Note that,
with a simple change of variable $u=2r^2/w^2$, and integrating by parts,
it is easy to verify that for any $z$
\begin{equation}
\label{(6)}
\int \!\!\! \int_{-\infty} ^{+\infty} {|\E|^2\over R}\,
\left(1-2{r^2\over w^2}\right) dx dy =0
\end{equation}
The net flux of reactive power through any plane orthogonal to the beam
axis is zero. Therefore, any ``space slice'' between such planes is
resonant, i.e. stores equal amounts of magnetic and electric energy.
But {\it locally}, this is not the case. Also note that these findings
match the previously outlined points on active power. Indeed, if on the
waist plane the wave impedance is smaller in the periphery compared
to the center, then the ratio of magnetic to electric energy has to
be larger in the periphery - and in fact it is. Finally, the interested
reader can calculate, through Maxwell's equations, the $\H_z$ component
of the magnetic field, and then the corresponding imaginary $y$-component
of the Poynting vector. Its sign reconfirms the previous statement:
electric energy stored around the axis is more than the magnetic one,
the opposite is true in the periphery.

\section{Analysis of free-space modes.}

One of the key issues of ref.\cite{MaSo87} was to show that a through-flowing
beam in free space can be correctly modeled, without running into
inconsistencies or paradoxes, as the superposition of two standing waves
of opposite parities. The typical example discussed in detail 
in ref.\cite{MaSo87} was a wave whose electric field, parallel to the
$z$-axis, is expressed, in cylindrical cordinates $r,\vartheta,z$
(see Fig.~4), as
\begin{equation}
            \E_z(r,\vartheta)=J_0(kr)+j J_1(kr)\sin(\vartheta)
   \label{campo}
\end{equation}
where $J_0,J_1$ are Bessel functions of the first kind of orders $0,1$,
respectively, and $k=2\pi/\lambda$ is the free-space wave number.
In ref.\cite{MaSo87}, no explicit statements were made on how the power
of this wave is distributed in space. However, it is legitimate to  
infer from the silence on this point, that it was taken for granted
in ref.\cite{MaSo87} what had been stated in ref.\cite{Mar85}, namely that 
the Poynting vector was everywhere proportional to the square of 
the modulus of the electric field, since free space is a 
homogeneous medium. In the previous Sections, we have shown that 
this attitude is not justified when dealing with Marcatili's tapers
or with Gaussian beams. In this Section we will prove that it is 
also erroneous in the case of free-space modes.

Let us first calculate the Poynting vector as $|\E|^2/2\eta_0$, where 
$\eta_0=\sqrt{\mu_0/\epsilon_0}$ is the free-space impedance. It is
straightforward to find that it has just a radial component 
expressed by
\begin{equation}
   \P_r={k\over2\omega\mu_0} \left[J_0^2(kr)+J_1^2(kr)\sin^2(\vartheta)\right]
   \label{Psb}
\end{equation}
This expression does not match with the idea of a through-flowing beam.
As a matter of fact, the flow lines for $\P_r$, depicted in Fig.~(5.a),
clearly give the feeling of an exploding wave, rather 
than of a through-flowing beam.

Let us now see what we find when we proceed rigorously, in the same
way as in the previous Sections. We calculate the magnetic field (two 
components) as the curl of eq.(\ref{campo}) 
in cylindrical coordinates, then the
Poynting vector (radial and azimuthal component). We take its real part,
and express it in terms of its cartesian components, in the reference 
frame shown in Fig.(4). These calculations yield:
\begin{equation}
\Re\{\P_x\}=-{k\over2\omega\mu_0}\left[
        J_0^2(kr)+J_1^2(kr)-2J_0(kr){J_1(kr)\over r}\sin(2\vartheta)
            \right]
   \label{repx}
\end{equation}
\begin{equation}
\Re\{\P_y\}=-{k\over\omega\mu_0}\left[
        \left(J_0^2(kr)+J_1^2(kr)\right)\sin^2(\vartheta)
        +2J_0(kr){J_1(kr)\over r}\cos(2\vartheta)
            \right]
   \label{repy}
\end{equation}
These results describe correctly a flow of active power, essentially 
in the direction of the $y$-axis. For example, to stress the fundamental
difference with respect to (\ref{Psb}), the fact that the sign of $\Re(\P_y)$
remains negative for $\vartheta=\pi/4,3\pi/4,5\pi/4,7\pi/4$ is 
perfectly adequate to describe a flow from the $y>0$ half space
towards the $y<0$ half space.

This is confirmed by Fig(5.b), where the flow lines for 
$\Re\{{\bf P}\}=\Re\{\P_x\}{\hat{\bf x}}+\Re\{\P_y\}{\hat{\bf y}}$ are
shown, and by Fig.~(6) which shows the space distribution of
the quantity $\Delta W=1/4(\mu_0|\H|^2 - \epsilon_0|\E|^2)$. We see that
the difference between magnetic and electric energy densities 
is far from being identically null. This confirms
that the field (\ref{campo}) is a {\it partially standing} wave. 
Once again, exactly like in the other cases discussed in the previous
Sections, the crucial difference between a standing wave
and a through-flowing beam is that the equal-amplitude loci of the
first one must not be confused with the surfaces orthogonal to
the Poynting vector of the second one.

\section{Conclusion.}

We tried to shed new light on an old problem, namely, whether the
idea of a guided mode traveling without any loss through a dielectric
taper can be sustained without running into any physical paradox.
Our numerical results, obtained with an extended BPM technique, have
fully reconfirmed what was stated in ref.\cite{MaSo87}: in Marcatili's tapers,
standing waves have the basic features outlined in ref.\cite{Mar85}, but
through-flowing waves do not. This prevents traveling waves from running
into a paradox, but on the other hand entails some loss radiation.
We have provided an explanation for the unexpected and
puzzling result, the drastic difference between standing and through-flowing
waves in the same structures. The source of these ``surprise'' is
built into Maxwell's equations.

It was pointed out in ref.\cite{MaSo87} that some of the problems discussed
here with reference to Marcatili's tapers apply to Gaussian beams
in free space as well. Indeed, in the rest of this paper we have
discussed Gaussian beams and free-space modes expressed in terms of Bessel
functions, and reached essentially the same conclusions as for
Marcatili's tapers.

\vspace{1cm}
\noindent
\section*{Acknowledgment} We gratefully acknowledge the contribution
given to the subject of Section~2 by Mr. Stefano Corrias, who passed away
in August 24, 1997.
\newpage

\section*{Figure captions}

\psn {\bf Fig.(1).} 
Constant-amplitude plots of two standing waves in a superlinear
Marcatili's taper, of even (part a)) and odd (part b)) symmetry, with
respect to the waist plane.

\psn {\bf Fig.(2).} 
Phase fronts (part a)), and field-amplitude contour plot (part b))
for a through-flowing wave in the same superlinear taper as in Fig.(1).

\psn {\bf Fig.(3).} 
Power vs. distance, in a superlinear taper of the shape shown in the previous
figures whose paramemters are specified in the text.

\psn {\bf Fig.(4).} 
Circular cylinder coordinate system.

\psn {\bf Fig.(5).} Flow lines for the real part of the Poynting vector
of free space modes. In inset a) the Poynting vector has been computed as
$\P=|{\bf E}|^2/2\eta_0$. Whereas, in inset b) it has been computed as
$\P={\bf E}\times{\bf H}^*/2$.

\psn {\bf Fig.(6).} 
Space distribution of the difference $\Delta W$ between magnetic and
electric energy densities.

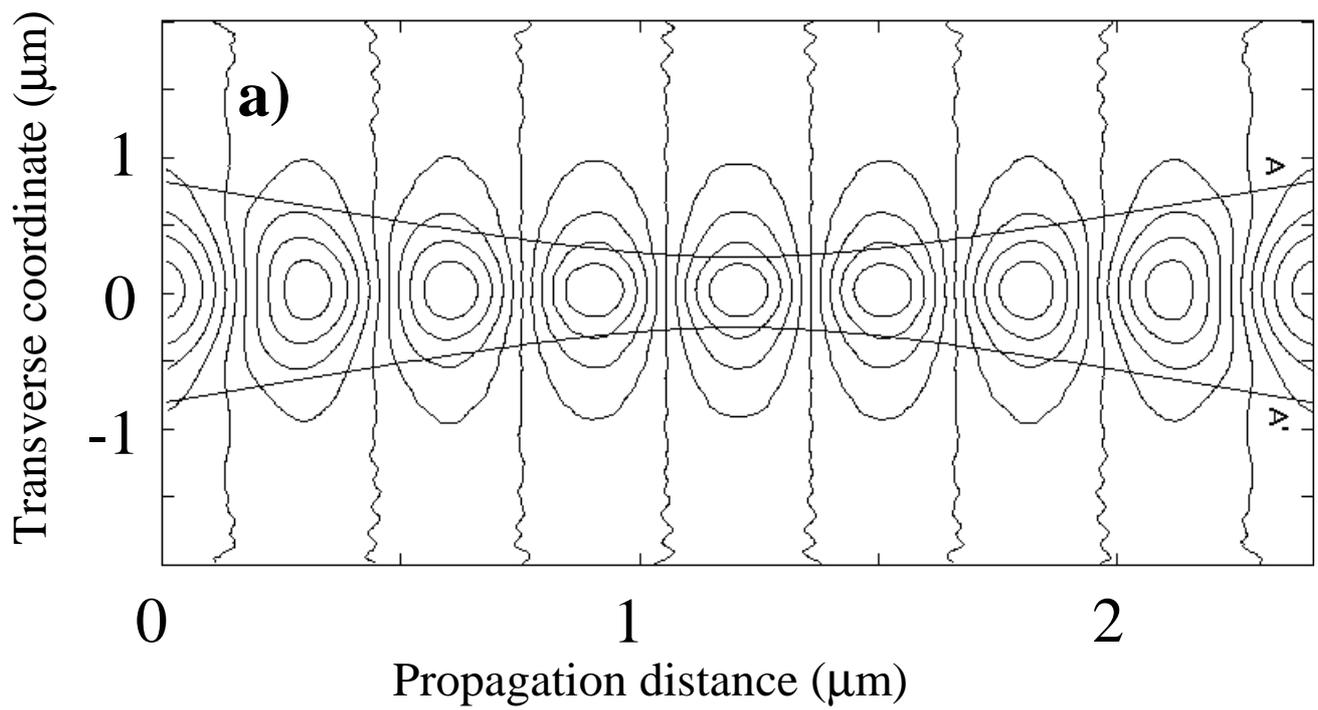
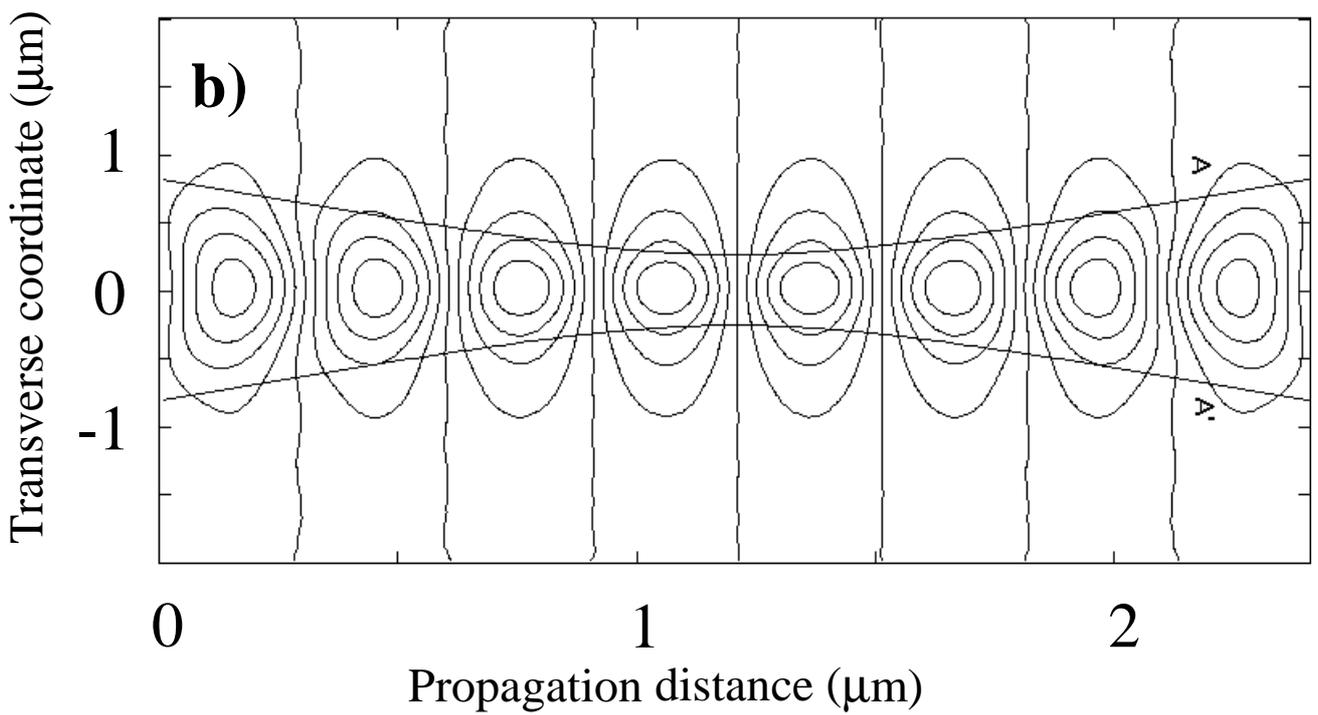
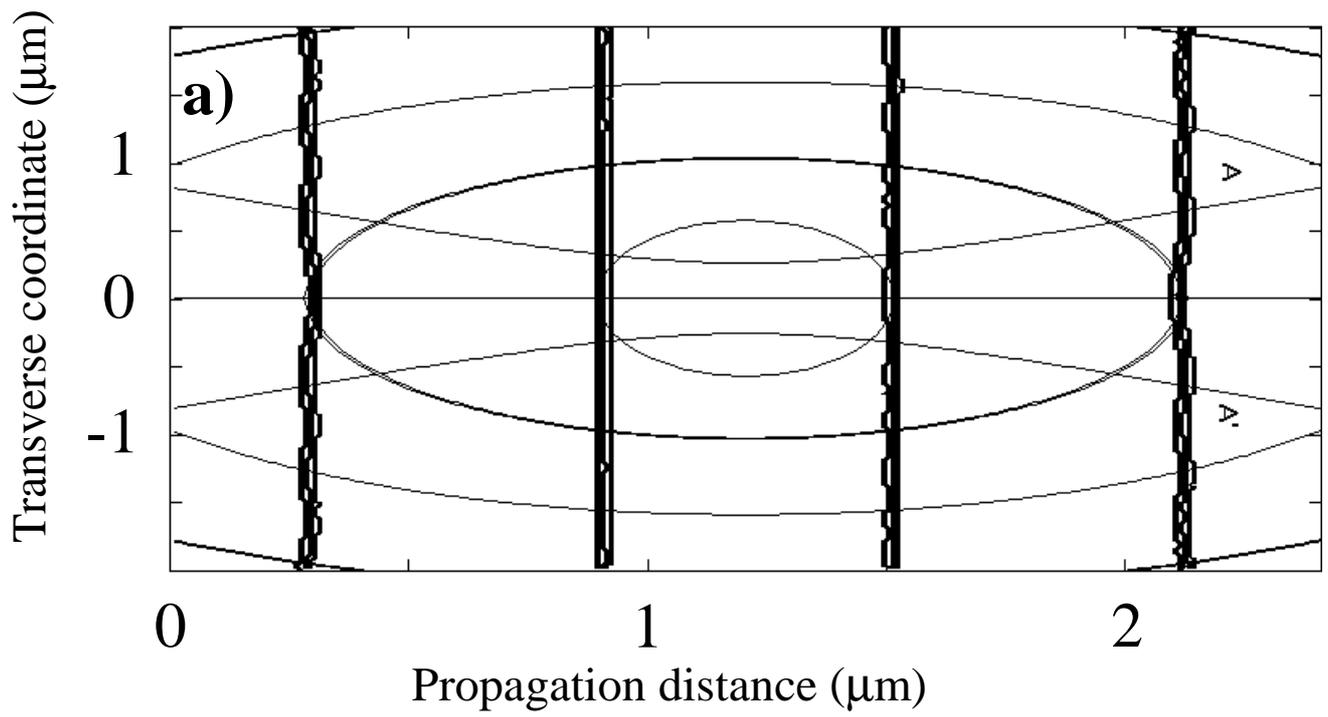
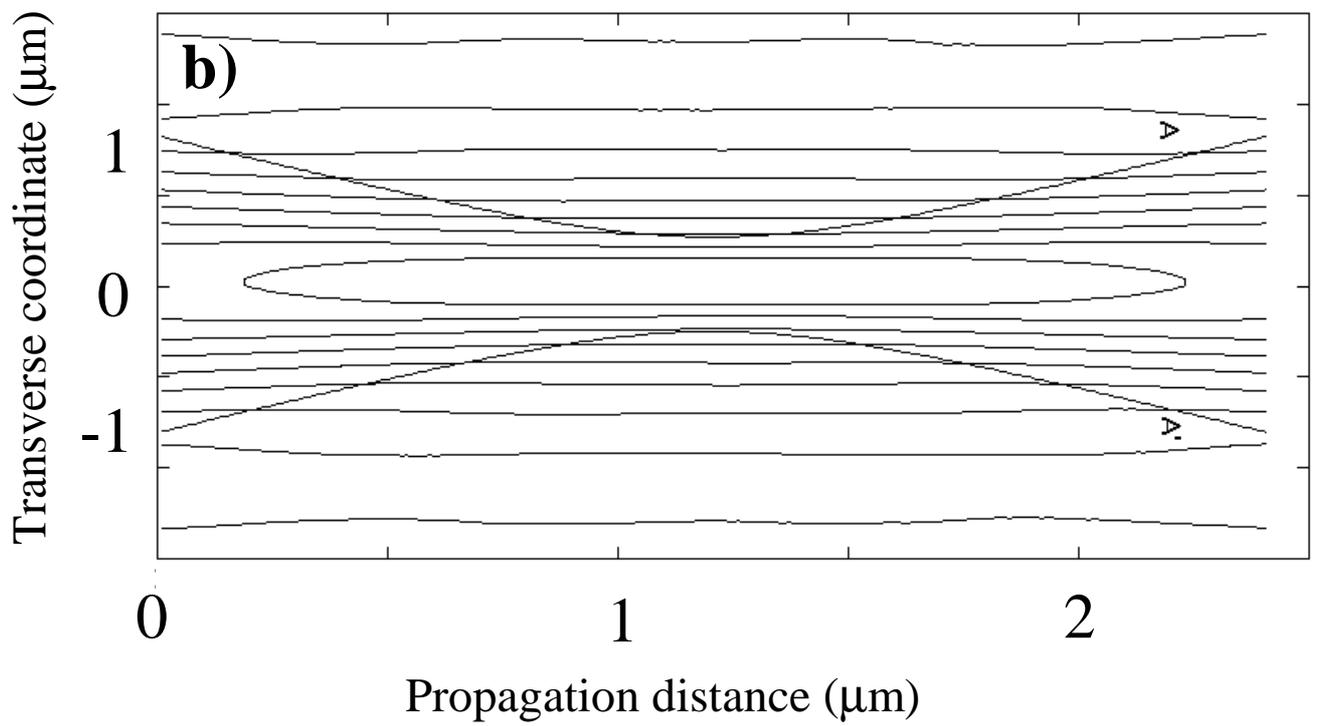
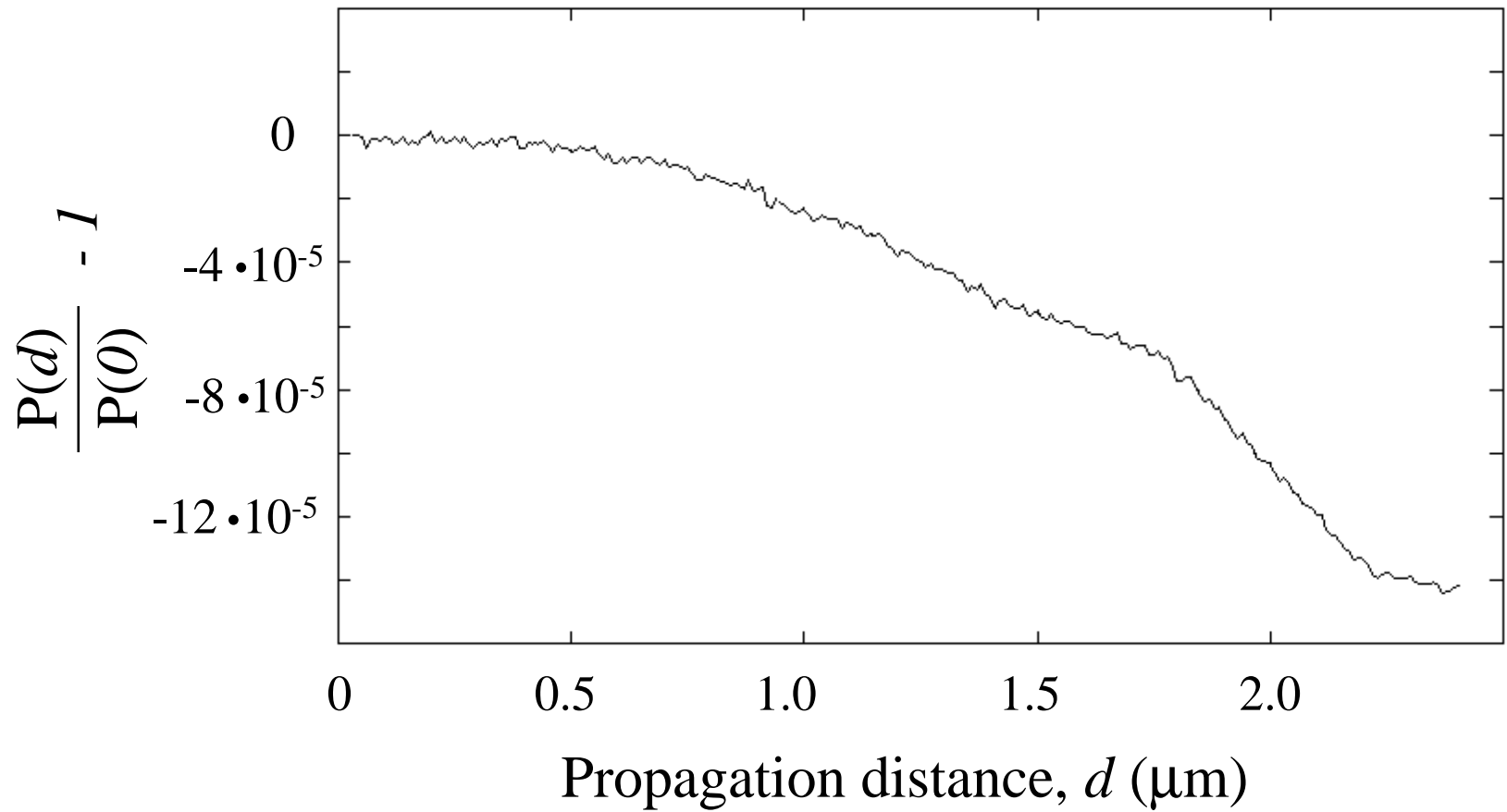
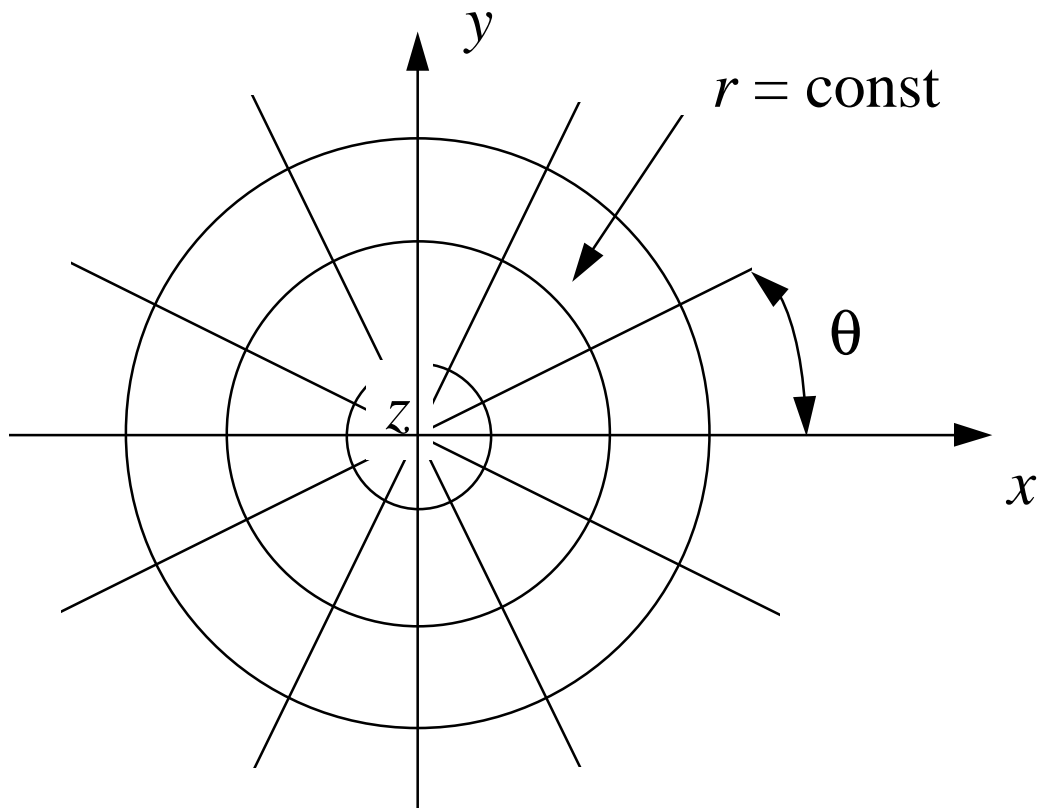
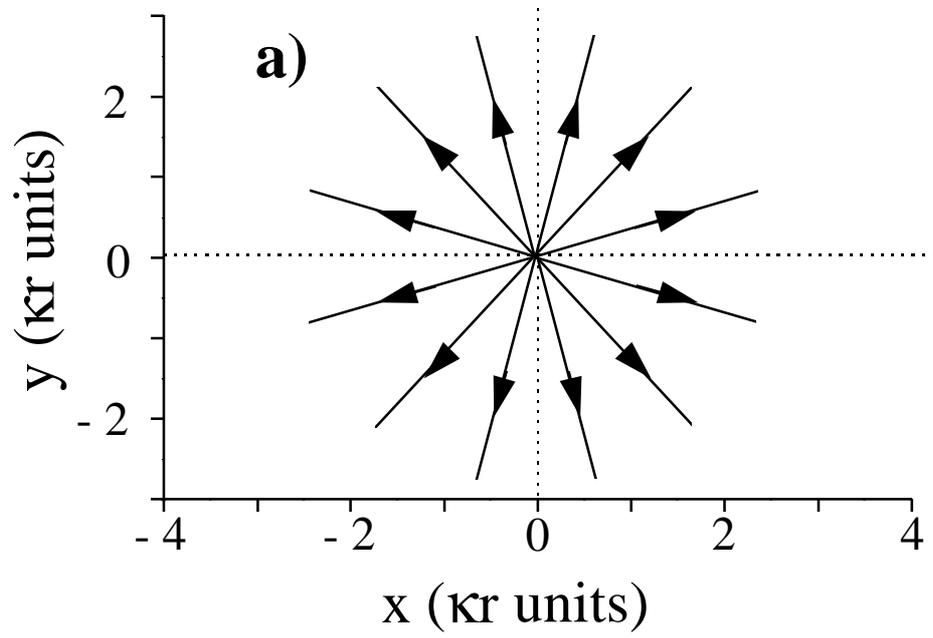
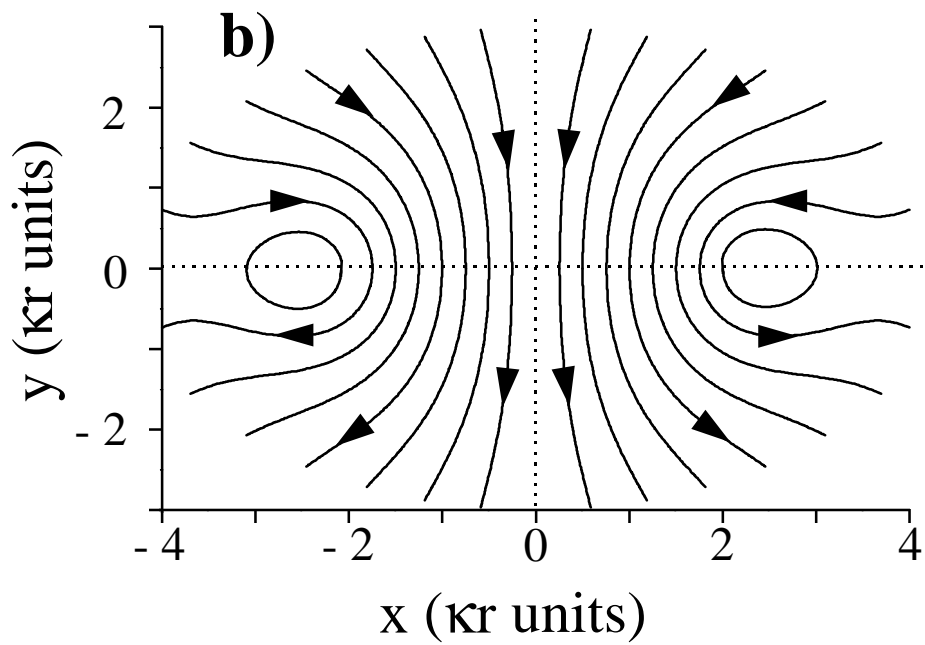
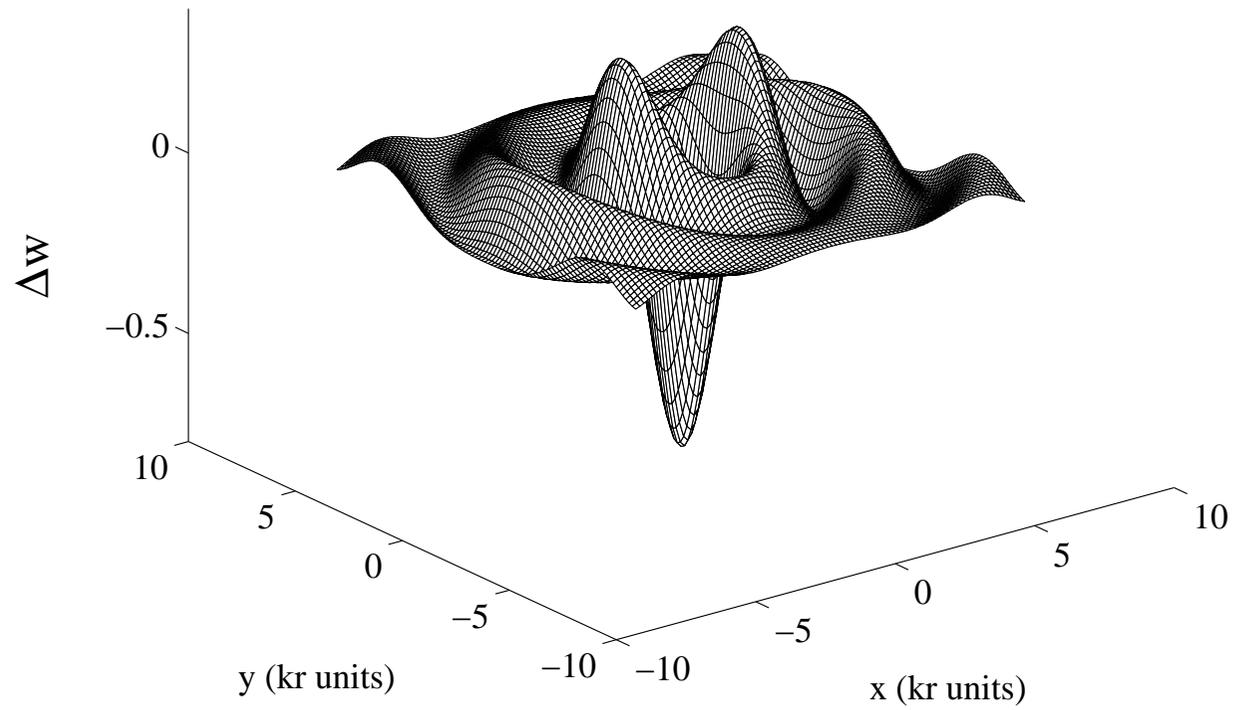
\end{document}